\documentclass{article}
\usepackage{spconf,amsmath,graphicx,epsfig,epstopdf}

\title{Full-info Training for Deep Speaker Feature Learning}
\name{Lantian Li, Zhiyuan Tang, Dong Wang, Thomas Fang Zheng\thanks{
This work was supported by the National Natural Science
Foundation of China under Grant No.61633013 / 61371136 and
the National Basic Research Program (973 Program) of China under
Grant No.2013CB329302. Lantian Li and Zhiyuan Tang are joint first authors.
Dong Wang is the corresponding author (wangdong99@mails.tsinghua.edu.cn).}}
\address{Center for Speech and Language Technologies, Research Institute of Information Technology \\
Department of Computer Science and Technology, Tsinghua University, Beijing, 100084, China}%
\begin{document}
%
\maketitle
\begin{abstract}

  In recent studies, it has shown that speaker patterns can be learned from very short
  speech segments (e.g., 0.3 seconds) by a carefully designed convolutional \& time-delay deep neural network (CT-DNN) model.
  By enforcing the model to discriminate the speakers in the training data, frame-level
  speaker features can be derived from the last hidden layer. In spite of its good performance,
  a potential problem of the present model is that it involves a parametric classifier, i.e.,
  the last affine layer, which may consume some discriminative knowledge,
  thus leading to `information leak' for the feature learning. This paper presents a full-info
  training approach that discards the parametric classifier and enforces all the discriminative
  knowledge learned by the feature net. Our experiments
  on the \emph{Fisher} database demonstrate that this new training scheme can produce more coherent features,
  leading to consistent and notable performance improvement on the speaker verification task.

\end{abstract}
\section{Introduction}

Automatic speaker verification (ASV) is an important biometric authentication technology and
has found a broad range of applications~\cite{beigi2011fundamentals,hansen2015speaker}.
The current ASV methods can be categorized
into two groups: the statistical model approach that has gained the most
popularity~\cite{Reynolds00,Kenny07,dehak2011front}, and the neural model
approach that emerged recently but has attracted much interest~\cite{ehsan14,heigold2016end,li2017deep}.

Perhaps the most famous statistical model is the Gaussian mixture model-universal background
model (GMM-UBM)~\cite{Reynolds00}. It factorizes the variance of speech signals
by the UBM, and then models individual speakers conditioned on that factorization.
Subsequent models design subspace structures to improve the statistical strength, including
the joint factor analysis approach~\cite{Kenny07} and the i-vector model~\cite{dehak2011front}.
Further improvements were obtained by either discriminative models (e.g., SVM~\cite{Campbell06}
and PLDA ~\cite{Ioffe06}) or phonetic knowledge
transfer (e.g., the DNN-based i-vector method~\cite{Kenny14,lei2014novel}).


The neural model approach has also been studied for many years~\cite{farrell1994speaker,Mueen2002Speaker},
however it was not as popular as the statistical model approach until recently
training large-scale neural models becomes feasible.
The primary success was reported by Ehsan et al. on a text-dependent
task~\cite{ehsan14}, where frame-level speaker features were extracted from the last hidden layer
of a deep neural network (DNN), and utterance-based speaker representations (`d-vectors') were derived by
averaging the frame-level features. Learning frame-level speaker features
is a key merit, which paves the way to deeper understanding of speech signals. This direction, however,
was not further investigated, as researchers quickly found that the relatively low performance of the d-vector approach
is due to the simple back-end, i.e., the average-based utterance representation. Therefore, many researchers
turn to seek for more complicated back-end models, e.g., Liu et al.~\cite{liu2015deep} used DNN
features to build conventional i-vector systems. Other researchers focused on the end-to-end approach that learned
utterance-level representations directly, e.g.,~\cite{heigold2016end,zhang2016end,snyder2016deep}.

In spite of the reasonable success of these `fat back-end' methods, we follow the feature learning
direction originated by Ehsan et al.~\cite{ehsan14}. Our assumption is that if speaker traits are short-time
identifiable and can be learned at the frame-level, many speech processing tasks will be much simpler, including ASV.
Fortunately, our recent study~\cite{li2017deep} showed that this frame-level speaker feature learning is feasible: with a short speech segment (0.3 seconds),
highly representative speaker features can be learned by a convolutional \& time-delay DNN (CT-DNN)
structure. Further study showed that these speaker
features are rather powerful: they can discriminate speakers by a short cough or laugh~\cite{zhang2017deep},
and work well in cross-lingual scenarios~\cite{li2017cross}. We also carefully compared the
feature learning approach and the end-to-end approach, and found that the feature learning
approach generally performed better, partly due to the more effective training scheme~\cite{wang2017deep}.

This paper follows the deep feature learning thread and extends our previous work in~\cite{li2017deep}.
The motivation is that the present CT-DNN architecture involves a parametric
classifier (i.e., the last affine layer) when training the feature learning component, or feature net.
This means that part of the knowledge involved in the training data is used to learn a classifier
that will be ultimately thrown away, leading to potential `information leak'.
This paper will present a full-info training approach that removes the parametric classifier
so enforces all the discriminative knowledge to be learned by the feature net.
Our experiments on the \emph{Fisher} database demonstrated that this new training scheme can
produce more coherent features, leading to notable and consistent performance improvement on the
speaker verification task.

In the next section, we will briefly describe the CT-DNN model that we proposed for
speaker feature learning, and then present the full-info training approach in Section~\ref{sec:method}. The
experiments are reported in Section~\ref{sec:exp}, and the paper is concluded in Section~\ref{sec:con}.

\section{Deep speaker feature learning}
\label{sec:speaker}

\begin{figure}[htb]
    \centering
    \includegraphics[width=0.98\linewidth]{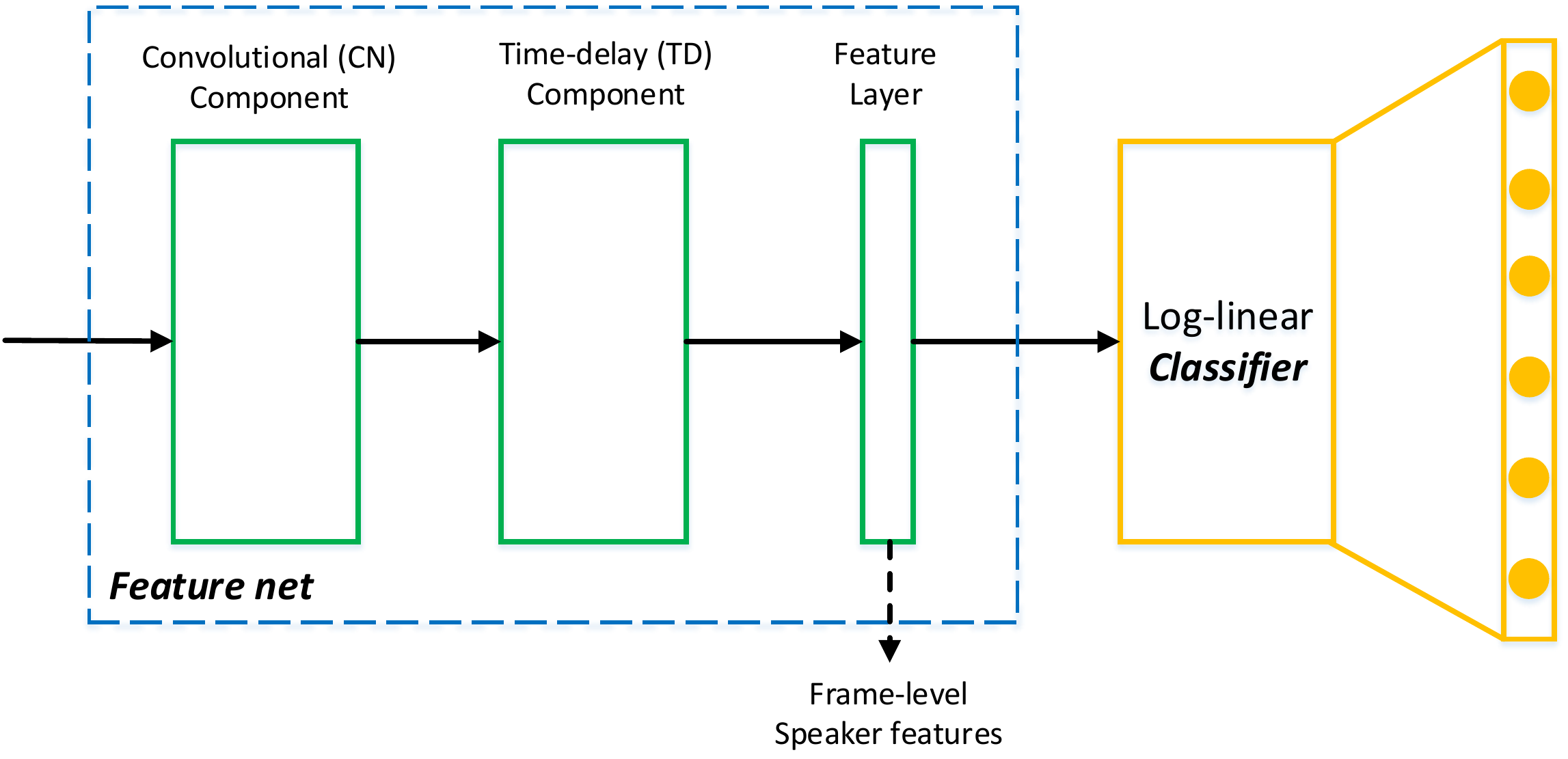}
    \vspace{-3mm}
    \caption{The CT-DNN structure used for deep speaker feature learning.}
    \label{fig:ctdnn}
\end{figure}

Fig.~\ref{fig:ctdnn} shows the CT-DNN structure that was presented in~\cite{li2017deep} and has been
demonstrated in several studies~\cite{zhang2017deep,li2017cross,wang2017deep}. The model consists of a convolutional (CN)
component and a time-delay (TD) component.
The output of the TD component is projected into a feature layer. The
activations of the units of this layer, after length normalization, form frame-level speaker features.
During model training, the feature layer is fully connected by an affine function to an output layer
whose units correspond to the speakers in the training data, which is essentially a classifier that
discriminates the target speakers in the training data on the basis of the input speech frame.
The training is performed to maximize the cross entropy of the classifier output and the ground truth label.
We have demonstrated that the speaker features inferred by the CT-DNN structure is
highly discriminative~\cite{li2017deep}, confirming our conjecture that speaker traits are largely short-time
spectral patterns and can be identified at the frame level.

\section{Full-info training}
\label{sec:method}

\subsection{Method}

The existing speaker feature learning models, either the vanilla structure proposed by Ehsan~\cite{ehsan14} or
our CT-DNN model~\cite{li2017deep}, involve two components: a \emph{feature net} and a \emph{classifier},
as shown in Fig.~\ref{fig:ctdnn}.
The feature net produces speaker-sensitive features, and the
classifier uses these features to discriminate the speakers in the training data.
In the CT-DNN case, the
features are produced from the last hidden layer, and the classifier is a log-linear model where
the non-linear activation function is softmax.
We emphasize that the feature net and the classifier are jointly trained. This is optimal if our task is to discriminate
the speakers in the training set, however when the features are used in other tasks, e.g., to discriminate or authenticate
other speakers, the joint training will be suboptimal. This is because the classifier involves \emph{free} parameters, so
part of the discriminant information will be learned by the classifier, which, unfortunately, will be thrown away
when performing identification/verification tasks on other speakers.

A possible solution is to discard the parametric classifier and using the speaker features to
classify the speakers directly. Specifically, if the frame-level
speaker features have been derived, each speaker $s$ in the training set can be represented by the average of
all the speaker features belonging to this speaker, given by:

\begin{equation}
\label{eq:vs}
 v(s; \theta) = \frac{1}{|\mathcal{E}(s)|} \sum_{x \in \mathcal{E}(s)} f(x; \theta),
\end{equation}

\noindent where $\mathcal{E}(s)$ is the set of speech frames belonging to speaker $s$, and $f(x; \theta)$ is the speaker feature
of frame $x$, produced by the feature net parameterized by $\theta$.
By these speaker vectors $v(s; \theta)$, each speech frame $x$ can be classified by a simple classifier as follows:

\begin{equation}
\label{eq:cos}
 p(s|f(x; \theta)) = \frac{e^{cos(f(x; \theta),v(s; \theta))}}{\sum_{s'} e^{cos(f(x; \theta),v(s'; \theta))}},
\end{equation}

\noindent where $cos(\cdot,\cdot)$ represents cosine distance.
The cross entropy between the classification output $p(s|f(x; \theta))$ and the ground truth label $s$ can be computed
and used as the cost function to train the system, formulated by:

\begin{equation}
\label{eq:cost}
  L(\theta) = \sum_{t} log \ p(s(t)|f(x(t); \theta))
\end{equation}
\noindent where $x(t)$ and $s(t)$ are the $t$-th speech frame and the corresponding ground truth label.
Note that cost function involves only $\theta$ as free parameters, so all the discriminative knowledge
provided by the training data is solely learned by the feature net. For this reason, we call this
approach \emph{full-info training}.

\subsection{Implementation}

\begin{figure}[htb]
    \centering
    \includegraphics[width=0.98\linewidth]{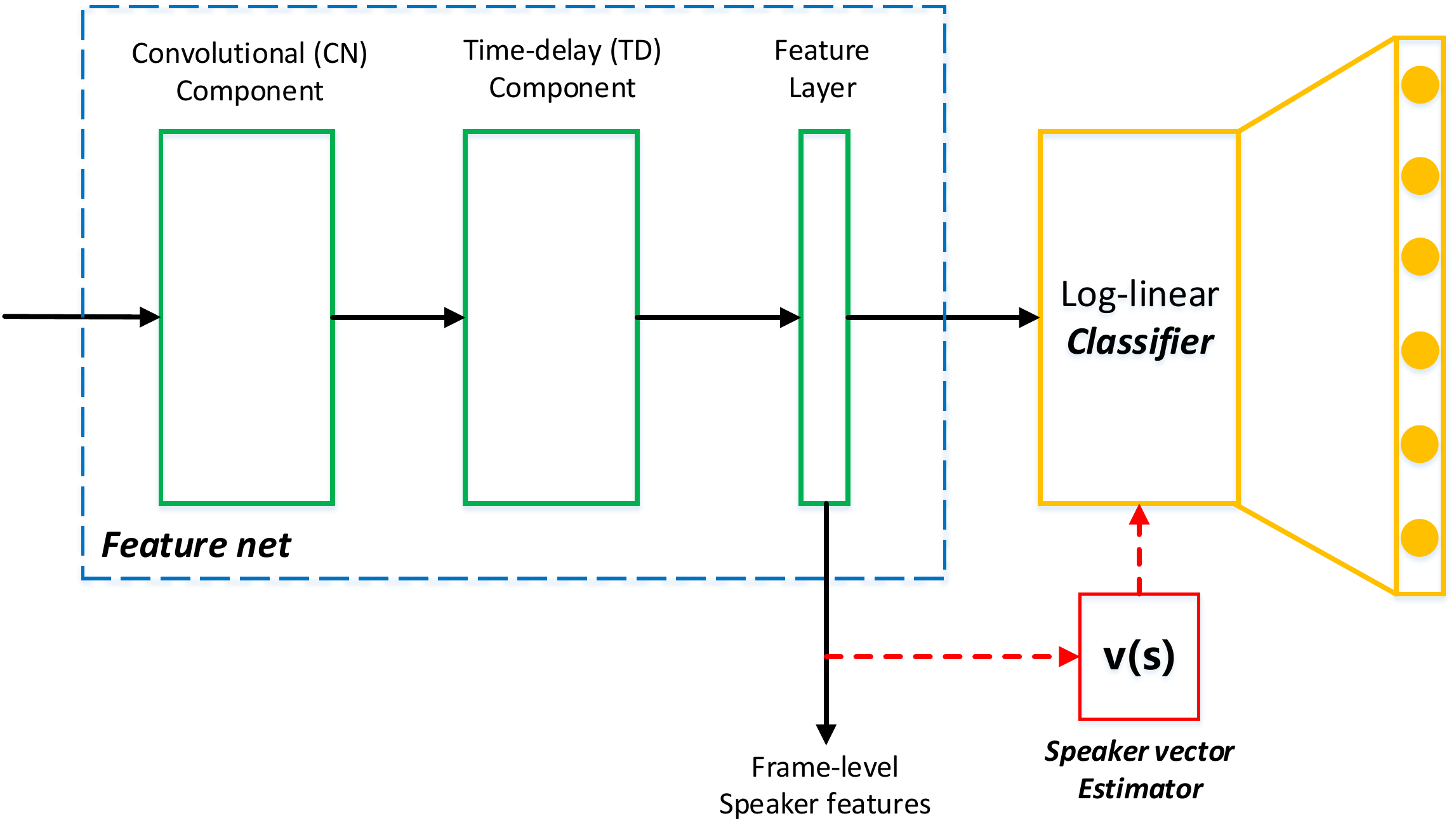}
    \caption{Iterative training scheme for full-info feature learning.}
    \label{fig:free}
\end{figure}

Optimizing Eq. (\ref{eq:cost}) is not simple, as $\theta$ appears in the speaker vectors $v(s;\theta)$. We design
an iterative scheme that can perform the optimization as a usual neural net training. As shown in Fig.~\ref{fig:free},
we keep the network structure (here CT-DNN) unchanged.
After each training epoch, the speaker vectors $v(s)$ are re-estimated according to Eq. (\ref{eq:vs}).
These speaker vectors are then used to replace the parameters of the log-linear classifier
(the last affine layer), and a new epoch is started following the regular back-propagation algorithm. Note that $v(s; \theta)$ should be
normalized to the same length when they are used to update the classifier, otherwise the forward computation of the last
affine layer (classifier) does not equal to Eq. (\ref{eq:cos}).

We experimented with various configurations to implement this iterative training, and found that
allowing the parameters of the classifier to be updated \emph{within} an epoch works slightly better
than keeping them fixed.
Another experience is that the a pre-trained
CT-DNN model can not be used as the initial for the full-info training; a random initialization for the feature net is required,
and the training is `warmed up' by using the speaker vectors produced by the pre-trained CT-DNN model.
Once this warm-up training converges, the iterative full-info can be started.


\subsection{Discussion}
\label{sec:method:dis}

The full-info training possesses several advantages: Firstly, all the discriminant knowledge
involved in the training data is learned from the feature net, so the training data is used more
effectively; Secondly, by the full-info training,
the frame-level features are encouraged to aggregate to their corresponding speaker vectors,
hence more coherent; Thirdly, the distance metric used in the full-info training (cosine distance)
is consistent with the measure used in the test phase of ASV. This improved
consistency is important when applying speaker features to the ASV task.

We note that the full-info training has been used in some end-to-end
approaches~\cite{heigold2016end,snyder2016deep}, though we focus on learning frame-level features
rather than utterance-level representations, and learning from multiple speakers
rather than speaker pairs as most of the end-to-end ASV approaches do.


\section{Experiment}
\label{sec:exp}

\subsection{Database}

The \emph{Fisher} database was used in our experiments. The training set and the evaluation set are presented as follows.

\begin{itemize}
    \item \textbf{Training set}: It consists of $2,500$ male and $2,500$ female speakers, with $95,167$ utterances randomly selected from the \emph{Fisher} database, and each speaker has about $120$ seconds speech segments. This dataset was used for training the UBM, T-matrix, and PLDA models
        of the i-vector system, and the CT-DNN model of the d-vector system.
    \item \textbf{Evaluation set}: It consists of $500$ male and $500$ female speakers randomly selected from
        the \emph{Fisher} database. There is no overlap between the speakers of the training set and the evaluation set.
        For each speaker, $10$ utterances are used for enrollment (about 30 seconds) and the rest are for test.

\end{itemize}

We test two scenarios: a short-duration scenario and a long-duration scenario.
Both scenarios involve $3$ test conditions.
For the short-duration scenario, the test conditions are `S(20f)', `S(50f)' and `S(100f)',
where the test utterances contain $20$, $50$ and $100$ frames respectively,
or equivalently $0.3$, $0.6$ and $1.1$ seconds.
For the long-duration scenario, the test conditions are `L(3s)', `L(9s)' and `L(18s)',
where the length of the test utterances is $3$, $9$ and $18$ seconds, respectively.

All the test scenarios/conditions involve pooled male- and female-dependent trials. Gender-dependent tests exhibit the
same trend, so we just report the results with the pooled trials.
Note that in the `S(20f)' condition, the length of the test utterances ($20$ frames)
is the size of the effective context window of the CT-DNN model, i.e., only one single
speaker feature can be derived.

\subsection{Settings}


We build two baseline systems: an i-vector system and a d-vector system based on the CT-DNN
structure. For the i-vector system, the feature involves $19$-dimensional MFCCs plus the log energy,
augmented by the first and second order derivatives. The UBM consists of $2,048$ Gaussian components,
and the dimensionality of the i-vector space is $400$. The entire system is trained
following the Kaldi SRE08 recipe. PLDA is used in scoring.

For the d-vector system, the raw feature involves $40$-dimensional Fbanks, and a symmetric $4$-frame
window is used to splice the neighboring frames.
The number of output units is $5,000$, corresponding to the number of speakers in
the training data. The speaker features are in $400$ dimensions, equal to the i-vectors.
The utterance-level d-vectors are derived by averaging the frame-level
speaker features. The Kaldi recipe to reproduce our results
has been published online\footnote{http://project.cslt.org}.
The scoring approach is the cosine distance on either the original $400$-dimensional d-vectors
or $150$-dimensional LDA-projected vectors.
Our previous experiments show that LDA can normalize the within-speaker
variation, which in turn normalizes the scores of different speakers. This is
important for speaker verification, as it is based on a global threshold and so requires
scores comparable across speakers.


\subsection{Main results}

The results in terms of equal error rate (EER\%) are reported in Table~\ref{tab:res}.
It can be observed that the d-vector baseline works better than the i-vector baseline in
the short-term conditions, but worse in long-term conditions. This tendency is
the same as in the previous studies~\cite{li2017deep,snyder2016deep}.

With the full-info training, we can observe that the performance of the d-vector system is
improved in a consistent way. Interestingly, in the short-term test conditions, the performance with
the cosine distance is not improved, but after LDA projection, the performance outperforms the
baseline. This indicates that the full-info training does not necessarily improve the
strength of single speaker features; instead, it encourages more coherent and generalizable
features. This is consistent with our discussion in Section~\ref{sec:method:dis}.

    \vspace{-2mm}
    \begin{table}[htb]
    \begin{center}
      \caption{EER results with different models and training methods on $6$ test conditions.}
      \label{tab:res}
      \scalebox{0.94}{
          \begin{tabular}{|l|l|c|c|c|}
            \hline
            \multicolumn{2}{|c|}{}                 &\multicolumn{3}{c|}{EER\%}\\
           \hline
               Models                 &  Scoring    &  S(20f) &  S(50f) &   S(100f)   \\
           \hline
               i-vector             
                                    &    PLDA     &  16.84    &   10.41      &   6.54      \\
            \hline
               d-vector             &    Cosine   &   7.89	  &   6.38       &    4.55      \\
                                    &    LDA      &   8.15    &   5.05       &    3.38      \\
            \hline
             d-vector               &    Cosine   &   9.48    &    7.45      &    4.74       \\
             + Full-info Training   &    LDA      &\textbf{7.53}&\textbf{4.36}&\textbf{2.85}  \\
            \hline
            \hline
              Models                &  Scoring    &   L(3s) &    L(9s)   &   L(18s)   \\
           \hline
               i-vector             
                                    &    PLDA     &   3.52    & \textbf{1.20}&   \textbf{0.89} \\
            \hline
               d-vector             &    Cosine   &   3.85    &    2.90      &    2.69      \\
                                    &    LDA      &   2.58    &  1.95        &    1.79      \\
            \hline
             d-vector               &    Cosine   &   3.95    &    2.48      &    2.23       \\
             + Full-info Training   &    LDA      &\textbf{2.14}  &  1.64      &    1.54      \\
           \hline

          \end{tabular}
          }
      \end{center}
   \end{table}
   \vspace{-8mm}

\subsection{Analysis}

\subsubsection{Training process}

\begin{figure}[htb]
    \centering
    \includegraphics[width=0.90\linewidth]{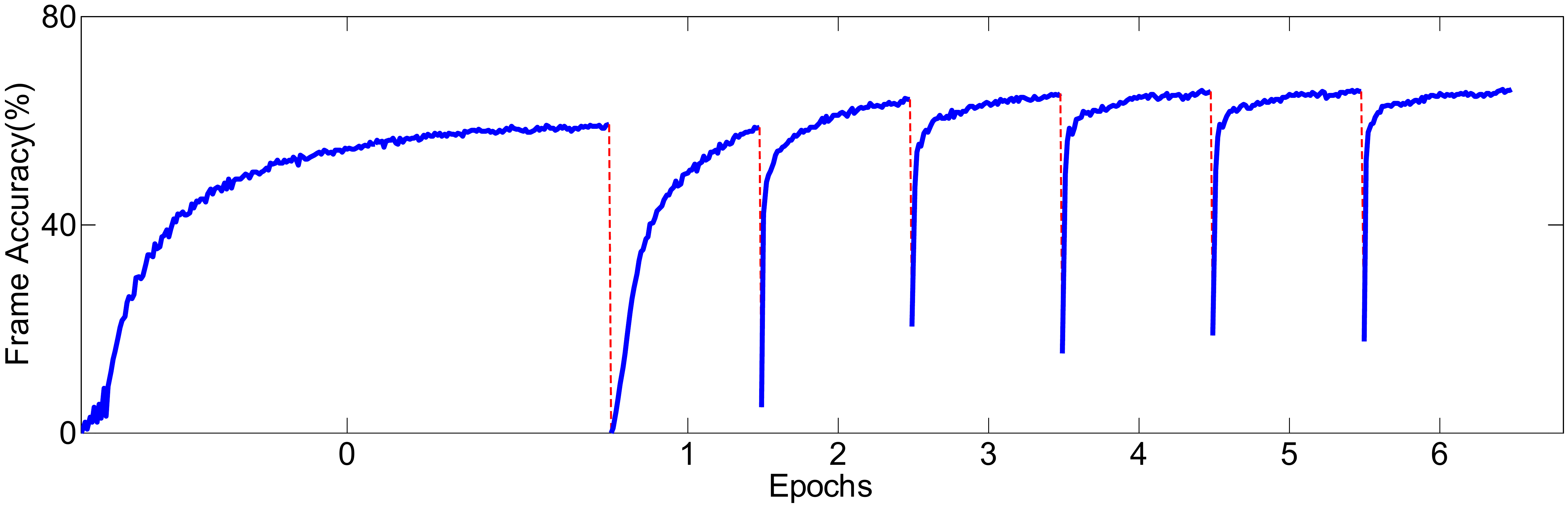}
    \vspace{-1mm}
    \caption{The change of frame accuracy on the validation set during the iterative full-info training.}
    \label{fig:training}
\end{figure}

Fig.~\ref{fig:training} presents the change of the validation-set frame accuracy during the iterative
training process (the trend on the training set is the same).
The epoch $0$ represents the basic CT-DNN model, which can be regarded as a `warm-up' model.
Once this warm-up training converges, the iterative full-info training will be started.

It can be observed that the accuracy is increased within each epoch, and after each
epoch, the initial accuracy starts from a higher value than the previous epoch. This indicates an increased
coherence between the frame-level speaker features and the speaker vectors. Note that the big gap between
the accuracies of the start and end stage of an epoch is due to the adaptable last affine layer.

\begin{figure}[htb]
    \centering
    \includegraphics[width=0.94\linewidth]{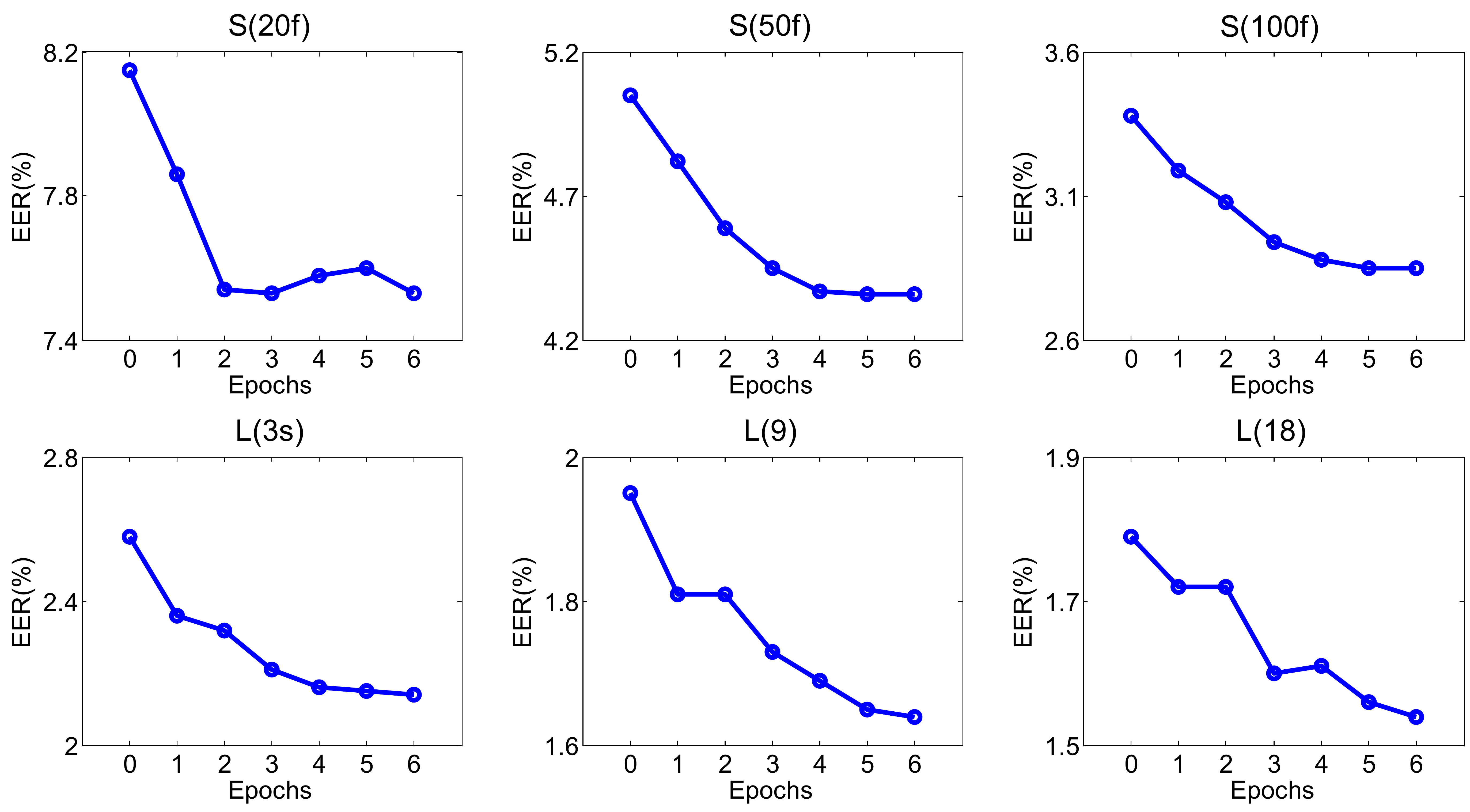}
    \caption{EER results on $6$ test conditions with iterative full-info training.}
    \label{fig:iterative}
\end{figure}

The EER results on the $6$ test conditions during the iterative training process
are shown in Fig.~\ref{fig:iterative}. We can observe a consistent and notable EER reduction
on all the test conditions.

\subsubsection{Visualization}

\begin{figure}[htb]
    \centering
    \includegraphics[width=0.96\linewidth]{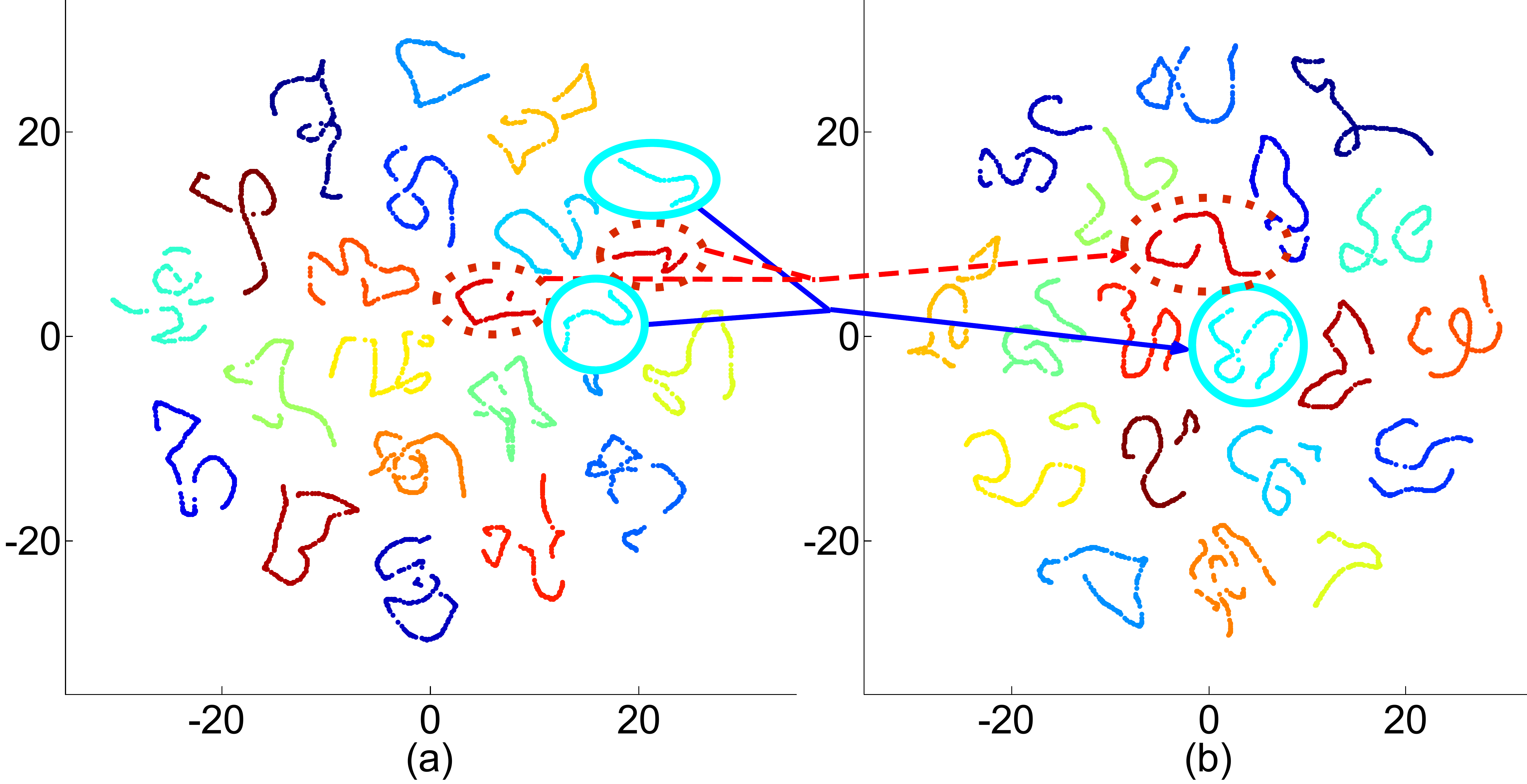}
    \caption{Deep speaker features are plotted by the t-SNE, with each color representing one speaker.
    Here (a) shows features produced by the original CT-DNN model, and (b) shows features produced by the CT-DNN model
    trained with full-info training.}
    \label{fig:continue}
\end{figure}

T-SNE~\cite{saaten2008} is used to visualize the speaker features in the 2-dimensional space.
In Fig.~\ref{fig:continue}, we choose several utterances from $20$ different speakers, and draw the speaker features
produced with and without the full-info training. It can be observed that the speaker features are highly discriminative,
no matter whether the full-info training is applied. However, the full-info training
produces more coherent features. Paying attention to the circled features, we observe that there
are two speakers (red and cyan) whose features are located in separated areas in the left
picture while aggregate together in the right picture. This clearly demonstrates that the full-info
training encourages more coherent features.

\section{Conclusions}
\label{sec:con}

This paper proposed a full-info training approach that
enforces all the speaker discrimination knowledge provided by the training data being
learned by the feature net, thus avoiding the `information leak' caused by the parametric
classifier involved in the conventional learning structure.
We tested the method on the speaker verification task with the Fisher database,
and found that it delivered consistent and notable performance improvement.
Methods that encourage more coherent features are under investigation.


\bibliographystyle{IEEEbib}
\bibliography{refs}

\end{document}